# Galactic chemical evolution with the short-lived isotopes 53Mn, 60Fe, 182Hf, and 244Pu

B. Wehmeyer[1,2,3,4,]*, A. Yagüe López[5], B. Côté[2,3,6], M. K. Pető[2,3], C. Kobayashi[4], and M. Lugaro[2,3,7,8]

[1]Institute of Theoretical Physics, University of Wrocław, 50-204 Wrocław, Poland

[2]Konkoly Observatory, HUN-REN Research Centre for Astronomy and Earth Sciences, Konkoly Thege Miklós út 15-17., H-1121, Hungary

[3]CSFK, MTA Centre of Excellence, Budapest, Konkoly Thege Miklós út 15-17., H-1121, Hungary

[4]Centre for Astrophysics Research, University of Hertfordshire, College Lane, Hatfield AL10 9AB, UK

[5]Computer, Computational and Statistical Sciences (CCS) Division, Center for Theoretical Astrophysics, Los Alamos National Laboratory, Los Alamos, NM 87545, USA

[6]Department of Physics and Astronomy, University of Victoria, BC, V8W 2Y2, Canada

[7]ELTE Eötvös Loránd University, Institute of Physics, Budapest 1117, Pázmány Péter sétány 1/A, Hungary

[8]School of Physics and Astronomy, Monash University, VIC 3800, Australia

**Abstract.** We run a three-dimensional Galactic chemical evolution (GCE) model to follow the propagation of $^{53}$Mn from supernovae of type Ia (SNIa), $^{60}$Fe from core-collapse supernovae (CCSNe), $^{182}$Hf from intermediate mass stars (IMSs), and $^{244}$Pu from neutron star mergers (NSMs) in the Galaxy. We compare the GCE of these short-lived radioactive isotopes (SLRs) to recent detections on the deep-sea floor. We find that although these SLRs originate from different sites, they often arrive conjointly on Earth.

## 1 Introduction

Galactic chemical evolution (e.g., [1–6]) is a valuable tool to study various astrophysical processes. Using short-lived ($\sim$Myr) radioactive isotopes (SLRs) provides additional timing information about these processes. We can use the exponential nature of radioactive decay to our advantage. If, for an SLR, out of the three values, (i) produced amount in an astrophysical nucleosynthesis site, (ii) observed amount, and (iii) the elapsed time between the production and the observation of the SLR, two values are known, conclusions can be drawn about the third one. This enabled, for example, determining the source, and production sites and conditions for various SLRs that influenced the early Solar system (cf., [7] for an extended review). There are multiple ways to observe the amount of an SLR:

1. By measuring the excess of the daughter isotope of the SLR compared to a reference isotope in meteorites to find their ratio at the time of the formation of the Solar system, to draw conclusions about the last nucleosynthesis processes before the formation of the Sun and which sites polluted the stellar nursery in which it was born (e.g., [7]).

---

*benjamin.wehmeyer@uwr.edu.pl

2. Measuring the γ-rays they emit during their decay (e.g., [8]) to draw conclusions about their distribution and production sites today (e.g., [9, 10]).

In this work, we analyse yet another way of measuring SLRs of astrophysical origin: $^{53}$Mn, $^{60}$Fe, and $^{244}$Pu of astrophysical origin were found in samples of the deep-sea floor (e.g., [11–14]). These SLRs rained down from the sky toward Earth over a time span of several Myr and slowly accumulated on the bottom of the ocean. Today, these can be found in deep-sea sediments and the Earth's crust. Analyzing these samples slice-by-slice, we can convert their occurrence rate in the slices to their density in the interstellar medium at the time represented by the depth of the slice (e.g., [11–14]). Here, we use these measurements to draw conclusions about the propagation of the three SLRs in the interstellar medium and also predict the presence of a fourth key SLR, $^{182}$Hf, in these samples.

## 2 Model

We used the cubic three-dimensional GCE model as in [15, 16] with an edge length of 2 kpc, periodic boundary conditions, and a sub-grid resolution of $(50\ pc)^2$. We then included the four SLRs of interest ($^{53}$Mn, $^{60}$Fe, $^{182}$Hf, and $^{244}$Pu) and their radioactive decay [17]. Below, we briefly summarize the calculations performed at every time step of 1 Myr:

1. Gas falls into the simulation volume according to a prescription that allows for an early rise, and late exponential decrease in infall.

2. We use a Schmidt law (with a power of α =1.5, [18–20]) to calculate the number of stars that are born during the current time step, and a Salpeter initial mass function with a slope of -2.35 ([21], with mass limits 0.1 $M_\odot$ ≤m≤50 $M_\odot$).

3. We use the Geneva group's [22–24] equation

$$\log(t) = (3.79 + 0.24Z) - (3.10 + 0.35Z)\log(M)$$

$$+ (0.74 + 0.11Z)\log^2(M), \tag{1}$$

to calculate the life expectation for every newly born star. t (in Myr) represents the expected lifetime of a star, Z is the metallicity, M represents the mass (in Solar masses) of the newly born star.

4. Most intermediate mass stars in the Galaxy are born in double- or triple star systems (e.g., [25]), hence there is a probability that this star is in a system that will later undergo a thermonuclear supernova explosion (SNIa). In order to omit the details of binary evolution, we use the probability $P_{SNIa} = 6 \cdot 10^{-3}$ as the fraction of all newly born IMSs to later undergo SNIa.

5. Correspondingly, there is a probability for a high mass star (HMS) to be born in a double star system. Comparably to SNIa, we reduce all details of binary evolution, and instead use a probability ($P_{NSM} = 0.04$) representing the fraction of all newly born HMSs to later end up in a NSM. The probability number can be converted into a gravitational wave emission rate (see [26] for details) ≈1, [27]).1800 Gpc₋₃ yr₋₁, slightly larger than the current limits of LIGO/Virgo (810 Gpc₋₃ yr-

6. Every time step corresponds to a list of stars whose life expectation is reached. Their death is simulated according to the details outlined below.

**2.1 Intermediate mass stars**

Once IMSs have reached the end of their life time, they eject their outer shells resulting in a planetary nebula. In our simulation, the star returns its mass as gas into the local sub-cell. Additionally, the dying star ejects $^{182}$Hf according to its mass. The yields are taken from Table S1 in the Supplementary Materials of [28].

**2.2 Massive stars**

Massive stars die as CCSNe. These explosions have a typical energy of the order of 1 Bethe. According to Sedov-Taylor shock wave theory, this corresponds to a swept-up gas mass of $5×10^4 M_\odot$. In our simulation, once a HMS has reached the end of its life time, it ejects alpha$^{60}$Fe according to yields in Table 3 in [31]. elements according to the yields in [29, 30], and

Then, the surrounding ISM is swept-up until $5×10^4 M_\odot$ of gas are displaced, forming a sphere around the CCSN progenitor.

**2.3 Thermonuclear supernovae**

When SNIa progenitor systems reach the end of their life time (i.e., the longer-lived IMS has died), comparable to CCSNe, the system ejects processed nuclei according to the yields in Table 3 in [32] (model CDD2), as well as $10^4 M_\odot$ of surrounding ISM is swept up by kinetic energy-$^4$ $M_\odot$ of $^{53}$Mn, in agreement with [33, 34].

Then, as in the CCSN case, $5 × 10$ deposition

into the ISM.

**2.4 Neutron star mergers**

In a NSM progenitor system, first, both HMS have to explode in a CCSN to produce the two neutron stars (NSs) required for an NSM. Once the two NS are formed, they orbit each other under the emission of gravitational waves, which reduces their angular momentum. In our simulation, we assume that after a coalescence time of $t_{coal} = 10^8$ years, the angular momentum of the system has decreased by enough for the two NSs to merge. Then, the NSM and total ejecta mass of 10ejects 10₋₈ $M_\odot$ of ₂₄₄Pu (in accordance with [35], assuming a mass fraction of X₋₂ $M_\odot$), and sweep up the surrounding gas as in the CCSN and₂₄₄ = 10₋₆,

SNIa case.

## 3 Results and Conclusions

Figure 1 shows the evolution of the four SLRs, zoomed in on 13.309 Gy $\lessapprox$ t $\lessapprox$ 13.332 Gy. We have opted to shift the times of the deposition of the SLRs on the deep-sea floor, to have a better fit of the simulated interstellar densities of the SLRs to the corresponding content of the SLRs in the deep-sea floor (we were more interested in the simulation following the shape of the deep-sea detections instead of the absolute values). In the Figure, we first look at the feature that is easiest to explain: radioactive decay. At t ≈ 13.330 Gy, the diagonal downward movement of the best-fitting density (green lines) seen in $_{60}$Fe and $^{182}$Hf stems from the radioactive decay of the isotopes. However, we realize strong changes in SLR densities that range over more orders of magnitude than the changes caused by radioactive decay (e.g., at 13.310 Gy in $^{53}$Mn, $^{182}$Hf, and $^{244}$Pu). These violent fluctuations are caused by nearby explosive events (CCSNe, SNIa, and NSMs) clearing the considered cell of its gas and SLR content. Further, we observe jump-like increases in different SLRs (e.g., at t = 13.314 Gy in $^{53}$Mn, $^{182}$Hf, and $^{244}$Pu). These can be explained by SLRs being transported together on a shock front of one or more explosive events: Consider a NSM exploding in a location just far enough from the local cell to not being influenced by that explosion. The shell around the explosive event (NSM) will be strongly enriched by $^{244}$Pu. However, we do not observe that increase in the sub-cell that we are currently looking at (green line in Figure 1). If now, inside the $^{244}$Pu enriched shell, a HMS explodes as a CCSN, $^{60}$Fe will be deposited on the $^{244}$Pu-enriched shell, and then shell will be pushed outwards from the former HMS location (due to the deposition of kinetic energy), carrying both the $^{60}$Fe, and the previously ejected (from the NSM) $^{244}$Pu with it. If the best-fitting cell (green line in Figure 1) happens to be in the location into which the explosion of the HMS pushes the remnant shell into, then both SLRs ($^{60}$Fe and $^{244}$Pu) are strongly and abruptly increased in that sub-cell. This explains, why in the best-fitting cell (green line in Figure 1), both SLRs can be increased simultaneously, although they originate in different nucleosynthesis sites. Although this case of a CCSN pushing the SLRs to adjacent cells is the most frequent case (since CCSNe are much more frequent then NSMs and SNIa), a similar argument can be made for all the other explosive sites. If an SLR (from one nucleosynthesis site) is present in one sub-cell, it can be transported to other sub-cells by an explosive event (CCSN, SNIa, or NSM), conjointly with the ejecta of that explosive event. In other words, we conclude that SLRs *surf the wave* of supernovae or NSMs.

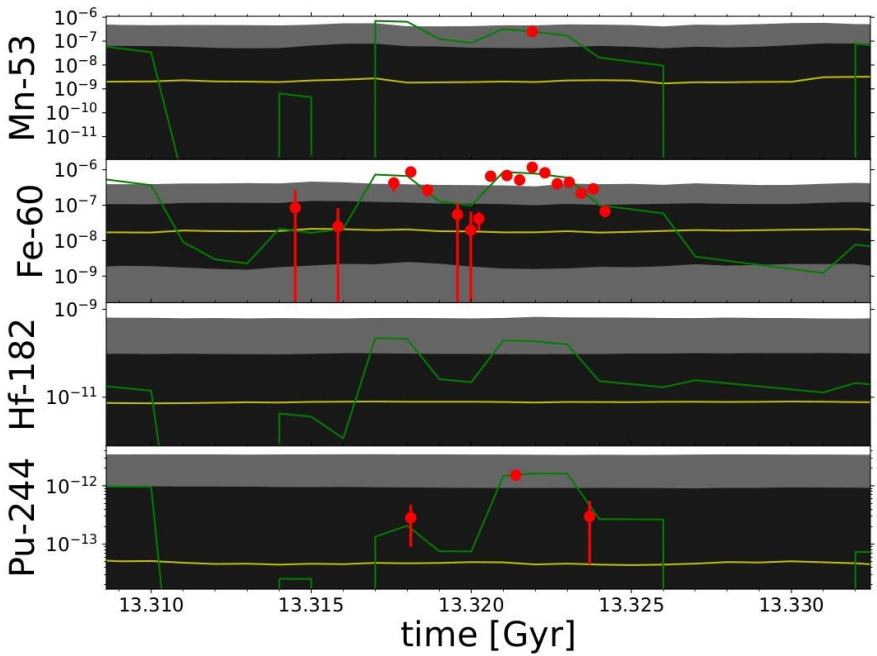

**Figure 1.** Evolution of the four SLRs ($^{53}$Mn, $^{60}$Fe, $^{182}$Hf, and $^{244}$Pu) in g/cm$^3$, zoomed-in on 13.309 Gy $\lessapprox$ t $\lessapprox$ 13.332 Gy. The yellow line represent the median density of the respective SLR in all sub-cells of the simulation volume, while the black (dark grey, light grey) shaded areas represent the 68% (95%, 100%) statistical fluctuations of the SLRs in all sub-cells. The deep-sea detections from [11, 12, 14] are represented as red boxes with error bars. The green line represents the evolution of the best-fitting sub-cell. We chose to shift the deep-sea detections on the x-axis, since we were more interested in the shape of the deep-sea detections, rather then their exact timing.

## Acknowledgements

BW acknowledges support from the National Center for Science (NCN, Poland), G. A. no. 2022/47/D/ST9/03092, as well as the National Science Foundation (NSF, USA), under G. A. no. PHY-1430152 (JINA Center for the Evolution of the Elements) and G. A. no. OISE1927130 (IReNA). CK acknowledges funding from the UK Science and Technology Facility Council (STFC) through G. A. no. ST/R000905/1, & ST/V000632/1. This work is supported by the ERC Consolidator Grant (Hungary) funding scheme (project RADIOSTAR, G.A. no. 724560). We also thank the COST actions "ChETEC" (G. A. no. 16117) and "ChETECINFRA" (G. A. no. 101008324). The work of AYL was supported by the US Department of Energy through the Los Alamos National Laboratory. Los Alamos National Laboratory is operated by Triad National Security, LLC, for the National Nuclear Security Administration of U.S. Department of Energy (Contract No. 89233218CNA000001). Some computations outlined in this paper were performed at the Wroclaw Centre for Scientific Computing and Networking (WCSS).